\documentclass{article}

\def\e{{\rm e}}\def\d{{\rm d}}

\usepackage{graphicx}\usepackage{epstopdf}\usepackage{bm}
\usepackage{amsmath}\usepackage{amssymb}

\begin{document}
\begin{center}
\textbf{Born-Infeld Black-Body Radiation}

\vspace{0.7cm}
Maryam Hajirahimi
$^{~(1}$
~~~and~~~Amir H. Fatollahi\footnote{email: fath@alzahra.ac.ir}
$^{(2}$
\vspace{0.7cm}

\textit{1) Physics Group,Islamic Azad University, South Tehran Branch, \\ P. O. Box 11365, Tehran 4435, Iran \\
\vspace{0.5cm}
2)  Department of Physics, Alzahra University, \\ P. O. Box  19938, Tehran  91167, Iran}
\end{center}

\vspace{0.7cm}

\begin{abstract}~
The problem of black-body radiation is considered in the Born-Infeld theory of
electrodynamics. In particular, at 2-loop order the deviation from the Planck
expression due to the self-interaction of photons is calculated. It is seen that the system of interacting photons of the theory, opposed to its non-Abelian counterpart, has higher internal energy at this order of perturbation. Possible implications of the result on the evolution of very hight temperature systems, including various stellar media and the early universe, are briefly discussed.
\end{abstract}
\vspace{1cm}

Pacs: 11.10.Wx, 
44.40.+a, 
14.70.Bh 


\newpage

The Born-Infeld theory of electrodynamics is given by the Lagrangian density \cite{bi}
\begin{equation}\label{1}
\mathcal{L}=-\sqrt{|\det(\gamma\, \eta_{\mu\nu}+F_{\mu\nu})|}
+\sqrt{-\det(\gamma\,\eta_{\mu\nu})}
\end{equation}
in which $F_{\mu\nu}=\partial_\mu A_\nu-\partial_\nu A_\mu$ and
$\eta_{\mu\nu}=\mathrm{diag}\,(1,-1,-1,-1)$ represents the space-time
metric. In the above $\gamma$ is a new constant with dimension (energy)$^{2}$ in the
units $\hbar=c=1$, by which one can recover the Lagrangian of Maxwell electrodynamics
in the limit $\gamma\to\infty$.
The original motivation to propose the non-linear theory above was to
overcome the problem of infinite field strength and energy
associated with the point charge source \cite{bi}. In particular,
the electrostatic field of a spherically symmetric solution of the
theory at the origin is finite and of order of $\gamma$.
It is shown that the Born-Infeld theory, though non-linear, possesses exact
plane-wave solutions \cite{boillat}.

The purpose of this letter is to study the problem of black-body radiation
in the electrodynamics theory of Born-Infeld. Due to non-linear character
of the theory we expect interaction among the photons.
Including lowest-loop corrections, we compute the deviation from the
black-body radiation of Maxwell theory. Possible implications of the result on the evolution of extremely- high-temperature systems, including various stellar media and the early universe, are briefly discussed.

In $1+3$ space-time dimensions the Lagrangian density (\ref{1}) reduces to
\begin{equation}\label{2}
\mathcal{L}= \gamma^2
\bigg(1-\sqrt{1+\frac{2Q}{\gamma^2}-\frac{P^2}{\gamma^4}} \bigg)
\end{equation}
in which $Q$ and $P$ are given by
\begin{align}\label{3}
Q&=\frac{1}{4}\, F_{\mu\nu}F^{\mu\nu} = \frac{1}{2}(\mathbf{B}^2 - \mathbf{E}^2) \\
\label{4}
P&=\frac{1}{4}\, {}^\ast F_{\mu\nu}F^{\mu\nu} =\mathbf{E}\cdot\mathbf{B}
\end{align}
with ${}^\ast F_{\mu\nu}=\varepsilon_{\mu\nu\rho\sigma}F^{\rho\sigma}/\,2$.
Evidently the Lagrangian is invariant under the Abelian gauge transformation
$A\to A+\partial \Lambda$, for any differentiable function $\Lambda(x)$.

The Lagrangian may be expanded in powers of $\gamma^{-1}$, for which, by
keeping the lowest orders, one gets
\begin{equation}\label{5}
\mathcal{L} = -\frac{1}{4}\, F_{\mu\nu}F^{\mu\nu} + \frac{1}{2 \gamma^2}
(Q^2+P^2) + O(\gamma^{-4})
\end{equation}
The first term above is simply the Maxwell term, by which the free theory is presented.
The other terms introduce interaction among the photons. By keeping the lowest order,
in the language of Feynman diagrams, we have a 4-leg vertex which is given by
\begin{align}\label{6}
 V_4=& \frac{1}{\gamma^2} \bigg[
 (p\cdot q) (r\cdot s)\, \eta^{\sigma\rho}\eta^{\mu\nu} +
(p\cdot r) (q\cdot s)\, \eta^{\mu\sigma}\eta^{\nu\rho} +
(p\cdot s) (q\cdot r)\, \eta^{\mu\rho}\eta^{\nu\sigma}  \cr
& +p^\nu q^\mu r^\rho s^\sigma + p^\sigma q^\rho r^\mu s^\nu + p^\rho q^\sigma r^\nu s^\mu
- (s\cdot r) p^\nu q^\mu \eta^{\sigma \rho} -(s\cdot p) q^\sigma r^\nu \eta^{\mu\rho} \cr
& - (s\cdot q) p^\sigma r^\mu \eta^{\nu\rho} -(q\cdot r) p^\rho s^\mu \eta^{\nu\sigma}
- (p\cdot r) q^\rho s^\nu \eta^{\mu\sigma} - (q\cdot p) r^\rho s^\sigma \eta^{\mu\nu}\cr
&+ (s^\alpha q^\beta \varepsilon_{\alpha\beta}^{~~~\rho\nu})(r^\kappa p^\lambda \varepsilon_{\kappa\lambda}^{~~~\sigma\mu})+
 (s^\alpha p^\beta \varepsilon_{\alpha\beta}^{~~~\rho\mu})(r^\kappa q^\lambda \varepsilon_{\kappa\lambda}^{~~~\sigma\nu})\cr
&+ (s^\alpha r^\beta \varepsilon_{\alpha\beta}^{~~~\rho\sigma})(q^\kappa p^\lambda \varepsilon_{\kappa\lambda}^{~~~\nu\mu})\bigg]
\end{align}
where the momenta and space-time indices are introduced in Fig.~1.
Also, the full quantum theory is given after adding gauge fixing term,
which may be taken as usual $-\frac{1}{2}(\partial\cdot A)^2$, and the corresponding
ghost sector. Since in the present theory the gauge group is Abelian, the ghosts
do not interact with the photons, hence the ghost sector simply can be integrated out.

\begin{figure}[t]\begin{center}
\includegraphics[scale=1]{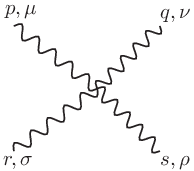}
\caption{The 4-leg photon vertex.}
\end{center}\end{figure}

Since the coupling constant $\gamma^{-2}$ has negative mass dimension,
the present theory, as a quantum field theory, is not renormalizable. This suggests
that the proper interpretation of the Born-Infeld theory would be as an
effective theory of an underlying finite or renormalizable theory.
In lack of the full underlying theory one still may try to explore the implications of the
Born-Infeld theory at higher orders of perturbation, for which one has to
give a recipe for the problem of infinities.
As mentioned, here we consider the black-body radiation problem, and the natural
framework for dealing with the interaction contribution
is the finite-temperature field theory \cite{fin-temp-book, kapusta-book}. As known, and
as we will see in more detail below, the problem of infinities finds a natural solution
in this framework \cite{fin-temp-book, kapusta-book}.
According to the recipe, the regulated finite-temperature
expressions are given simply after the subtraction of the
zero-temperature counterpart expressions \cite{kapusta-book}. By this,
the study of thermodynamical properties may give a basis to explore
the implications of the effective theory at higher orders of perturbations.

At 1-loop order (no vertex), the free Maxwell theory and the contribution
from the ghost sector simply give the standard Planck expression
for black-body radiation. The first contribution from interactions
comes at 2-loop (one vertex) order. As mentioned, to get the relevant
regulated contributions one subtracts an altered form of the
diagrams, the so-called parenthesized diagrams, from the original
ones. At 2-loop order by ``parenthesized" we simply mean that in
one of the loops we send temperature to zero
\cite{kapusta-book,kapusta}. In the present case we have the
contribution from the diagrams in Fig.~2 \cite{kapusta-book,kapusta}.

\begin{figure}[h]\begin{center}
\includegraphics[scale=1]{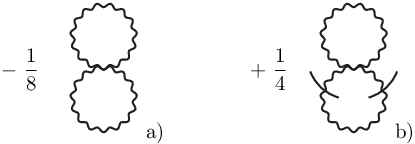}
\caption{a) the original 2-loop diagram. b) the parenthesized version of a).}
\end{center}\end{figure}

Using the identity
\begin{equation}\label{7}
\varepsilon^{\mu\nu\rho\sigma} \varepsilon_{\mu\nu}^{~~~\kappa\lambda}
= -2(\eta^{\rho\kappa}\eta^{\sigma\lambda}-\eta^{\rho\lambda}\eta^{\sigma\kappa})
\end{equation}
the original diagram (Fig.~2a), in units $c=\hbar=k_\mathrm{B}=1$, is associated with the expression
\begin{align}\label{8}
\Delta \mathcal{F}_\mathrm{2-loop}^\mathrm{orig.}=\frac{-1}{8\gamma^2}
\int \frac{\d^3p}{(2\pi)^3}\int\frac{\d^3q}{(2\pi)^3}\; \frac{1}{\beta} \sum_{n_p} \frac{1}{\beta} \sum_{n_q}
\bigg(\frac{8(p\cdot q)^2}{p^2 q^2}+7\bigg)
\end{align}
in which $\beta=T^{-1}$ is the inverse of the temperature, and the 4-vectors $p$ and $q$ have the general
form of $k=(2\pi i n_k T, \mathbf{k})$, with $n_k$ taking all integer values. In the parenthesized form
of the diagram one takes the limit $\beta\to \infty$ in one of the summations of the above expression,
turning the integer variable to a continuous one \cite{kapusta}.
Following \cite{kapusta}, one can transform the summations to integrals in the complex plane
\begin{align}\label{9}
T\sum_{n=-\infty}^\infty f(k_0=2\pi i n  T)  = & \frac{1}{2\pi i} \int_{-i\infty +\epsilon}^{i\infty +\epsilon}
\d k_0 \frac{1}{\exp(\beta k_0)-1} \big[f(k_0) + f(-k_0)] \cr
& + \frac{1}{2\pi i} \int_{-i\infty}^{i\infty} \d k_0 f(k_0), ~~~~~~~\mathrm{with}~~\epsilon\to 0^+.
\end{align}
We mention that the second term is independent of temperature, and survives in the limit $\beta\to\infty$.
The first term can be calculated by the residues at the poles of the function $f(k_0)$ in the right half-plane.

By using the above one finds for the original diagram
\begin{align}\label{10}
\Delta\mathcal{F}_\mathrm{2-loop}^\mathrm{orig.}=\frac{-1}{8\gamma^2}& \int \frac{\d^3p}{(2\pi)^3}\int\frac{\d^3q}{(2\pi)^3}\cr
&\bigg\{\frac{4N_p N_q} {\omega_p \omega_q} \Big[(\omega_p\omega_q - \mathbf{p}\cdot\mathbf{q})^2
+ (\omega_p\omega_q + \mathbf{p}\cdot\mathbf{q})^2 \Big] \cr
&+2\frac{N_p+ N_q} {\omega_p \omega_q} \Big[(\omega_p\omega_q - \mathbf{p}\cdot\mathbf{q})^2
+ (\omega_p\omega_q + \mathbf{p}\cdot\mathbf{q})^2 \Big] \bigg\}\cr
&+ \infty\mathrm{~terms~independent~of~}T
\end{align}
in which $N_k=\big(\exp(\beta \omega_k) -1\big)^{-1}$, with $\omega_k=|\mathbf{k}|$.
After subtracting the parenthesized diagram, the contribution from 2-loop order to the
free energy per unit volume is found
\begin{align}\label{11}
\Delta\mathcal{F}_\mathrm{2-loop} = & \frac{-1}{\gamma^2} \int
\frac{\d^3p}{(2\pi)^3}\int\frac{\d^3q}{(2\pi)^3}\; N_p N_q
\frac{\omega_p^2\omega_q^2 +
(\mathbf{p}\cdot\mathbf{q})^2}{\omega_p \omega_q}\cr
& + \infty\mathrm{~terms~independent~of~}T.
\end{align}
Using
\begin{equation}\label{12}
\int_0^\infty\frac{s^{2m +1}\d s}{\e^{s}-1}=(2m +1)!\;\zeta(2m +2)
\end{equation}
with $\zeta(t)$ as the Riemann zeta function, one can do the integrations over
$\d^3q=\omega_q^2\d\omega_q\d\Omega_q$ and $\d\Omega_p$ to get
\begin{equation}\label{13}
\Delta\mathcal{F}_\mathrm{2-loop} = -\frac{T^4}{45\,\gamma^2}
\int_0^\infty \frac{\omega^3 \d\omega}{ \exp(\beta\omega)-1 }
\end{equation}
The internal energy per unit volume is then given by
\begin{align}\label{14}
U&=\mathcal{F} +\beta \partial_\beta \mathcal{F} \\
\label{15}
&=\int_0^\infty\d\omega ~U(\omega)
\end{align}
for which we find
\begin{equation}\label{16}
\Delta U_\mathrm{2-loop}(\omega) =  \frac{T^4 }{45\,\gamma^2}
\omega^3 \left[  \frac{3 }{\exp(\beta \omega)-1}
+ \frac{\beta\omega \exp(\beta\omega)}{(\exp(\beta \omega)-1)^2}\right]
\end{equation}
as the energy per unit volume in the frequency interval $(\omega,\omega+\d\omega)$.
In fact this expression gives the correction to the standard Planck expression
\begin{equation}\label{17}
U_\mathrm{Planck} (\omega)= \frac{1}{\pi^2}\frac{\omega^3}{\exp(\beta\omega)-1},
\end{equation}
by which the Stefan's law follows,
\begin{equation}\label{18}
U_\mathrm{Planck}=\int_0^\infty\d\omega ~U_\mathrm{Planck}(\omega)=\frac{\pi^2}{15}\; T^4.
\end{equation}
Using (\ref{12}) and
\begin{equation}\label{19}
\int_0^\infty\frac{s^{2m +2}\e^{s}\d s}{(\e^{s}-1)^2}=(2m +2)!\;\zeta(2m +2)
\end{equation}
one finds the following as the correction to Stefan's law
\begin{equation}\label{20}
\Delta U_\mathrm{2-loop}=\frac{7\pi^4}{675} \;\frac{ T^8}{ \gamma^2}.
\end{equation}
So the total energy is found to be
\begin{equation}\label{21}
U_\mathrm{Born-Infeld}=\frac{\pi^2}{15}\; T^4+ \frac{7\pi^4}{675} \;\frac{ T^8}{ \gamma^2}.
\end{equation}
This expression might be contrasted to the analogues one in non-Abelian gauge theories.
For SU($N_c$) pure gauge theory, the combination of the free theory and first correction
is found to be \cite{kapusta}
\begin{equation}\label{22}
U_{\mathrm{pure~SU}(N)}=\frac{\pi^2}{15}\;N_\mathrm{g}\; T^4- \frac{\pi}{3} \;\alpha_c\;N_c N_\mathrm{g} T^4,
\end{equation}
in which $\alpha_c=g^2/(16\pi)$ is the color fine-structure constant, and $N_\mathrm{g}=N_c^2-1$ is the number of the gluons of the theory. Most importantly, due to the minus sign, taking into account the interactions between the gauge bosons would decrease the internal energy and so the pressure in the systems, as befits the name ``gluon" for the gauge bosons in the non-Abelian case. In comparison,
the photons of the Born-Infeld theory do not exhibit the glue behavior, and in fact act in opposite way.

\begin{figure}[t]\begin{center}
\includegraphics[scale=0.4]{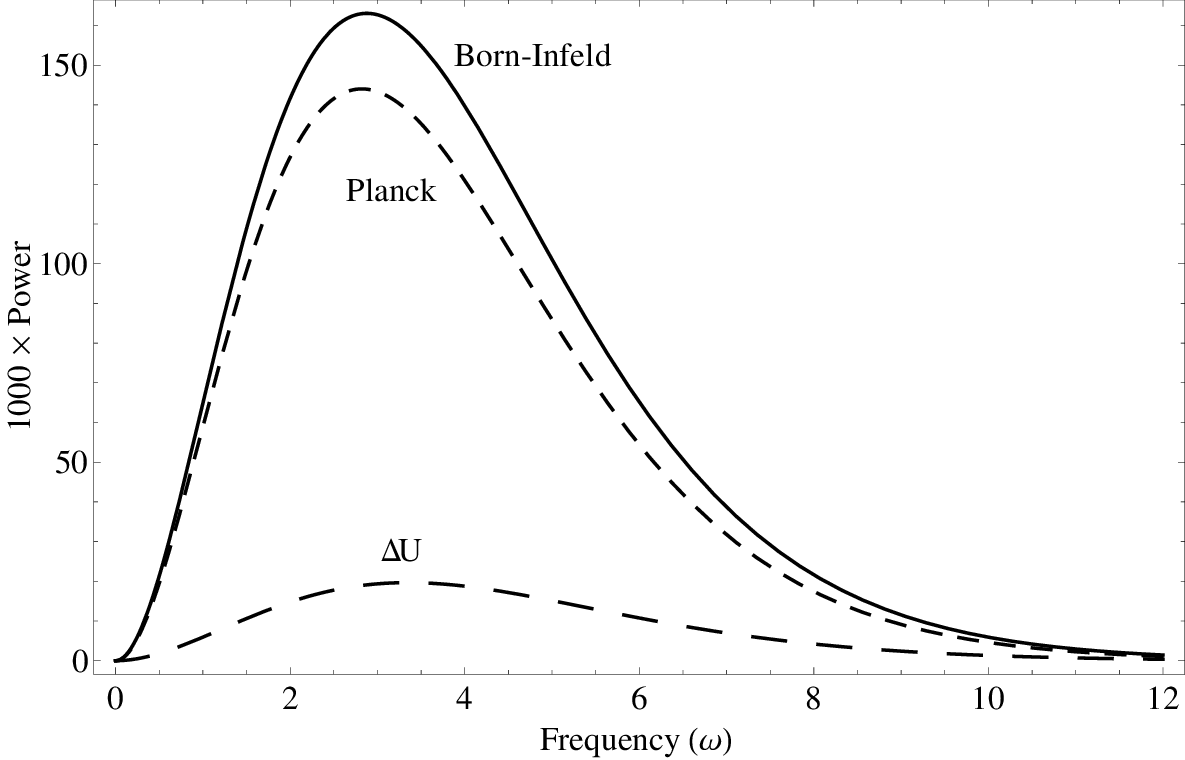}
\caption{Plots of internal energy versus frequency, for $\beta=1$ and $\gamma^2=10$.
The solid line is the sum of Planck expression and $\Delta U(\omega)$ of (\ref{16}).}
\end{center}\end{figure}

To get a better idea of the kind of the changes by including the 2-loop correction, in Fig.~3
the plots of the Planck expression, $\Delta U(\omega)$ of (\ref{16}), and their combination
are presented.

The test of the Born-Infeld theory has been the subject of several researches, among them are the tests based on, the waveguide physics \cite{fer1}, and explored the consequences of the non-linearity of the theory in the presence of background fields on the induced anisotropy of the electromagnetic waves \cite{fer2}, the change in the measured luminosity of distant objects \cite{fer3}, and the wave dispersion relation of a propagating laser in a ring \cite{deni}.

One may try to get an idea of the magnitude of the possible deviation by the present results from the classical expressions. There exists lower limits on the parameter $\gamma$ in the theory. As mentioned $\gamma$ plays the role of the maximum
possible value for the electric field, the so-called absolute field constant \cite{bi}.
Based on the electric-field value at the edge of the classical radius of electron,
Born and Infeld estimated $3\times 10^{20}$~V/m
in SI units as the lower limit of the absolute constant field, which corresponds to $\gamma \gtrsim4\times 10^{14}$~eV$^2$, in the
natural units. The order of this lower limit is supported by
investigations done on the atomic energy levels \cite{jackson}. Having the lower limit for $\gamma$, there are several ways to have an idea of the upper limit of possible deviations. One is the change in the total radiation energy. By (\ref{18}) and (\ref{20}),
and after restoring the Boltzmann constant $k_\mathrm{\,B}=8.6\times 10^{-5}$~eV/K, by the above lower limit for $\gamma$ one finds
\begin{equation}\label{23}
\frac{\Delta U_\mathrm{2-loop}}{U_\mathrm{Planck}} =  \frac{7\pi^2}{45}
\frac{(k_\mathrm{B} T)^4}{\gamma^2}
\lesssim 5.2\times 10^{-46}\,T^4.
\end{equation}
A very close estimation is obtained for the ratio at the maximum frequency,
for which one has by the Wien's law $\beta \,\omega_\mathrm{max}\simeq 2.8$.
With $\gamma$ equal to the lower limit and the temperatures higher than $10^{11}$~K  the above ratio finds values that are expected to have remarkable implications on the models, as well as possible observational indications. By the present models for the neutron stars, the cores of young neutron stars have the temperatures in the range $10^{10-12}$~K. It should be reminded, due to perturbative nature of the presented calculation, in the cases where the ratio (\ref{23}) exceeds unity, the analysis can give only a qualitative description. As the energy distribution of the photons inside a star has direct implications on the evolution of star, it is reasonable to look for the possible consequences of the present results.

Another possibility is to check the ratio of the two distributions (\ref{16}) and (\ref{17}) at their tails, where $\beta\omega\gg 1$. After restoring the Planck constant $\hbar$
one gets
\begin{equation}\label{24}
\left.\frac{\Delta U_\mathrm{2-loop}(\omega)}{U_\mathrm{Planck}(\omega)}\right|_{\beta\omega\gg 1} \simeq  \frac{\pi^2}{45}
\frac{(k_\mathrm{B} T)^3}{\gamma^2}\cdot\hbar\omega \lesssim
8.7\times 10^{-43}\, T^3\cdot \hbar\omega
\end{equation}
in which the last value is again obtained for $\gamma$ equal to its lower limit value.
In Fig.~4 the increasing ratio (\ref{24}) of the number of photons versus the photon energy for three different temperatures
are presented. Here we see that there is a variety of internal star media, including the cores of high-mass stars, that can give an increase from few percents to few times in the number of their high-energy photons. Again it is reminded that the ratios bigger than unity can only give a qualitative description of the behavior of the present theory.

\begin{figure}[t]\begin{center}
\includegraphics[scale=0.4]{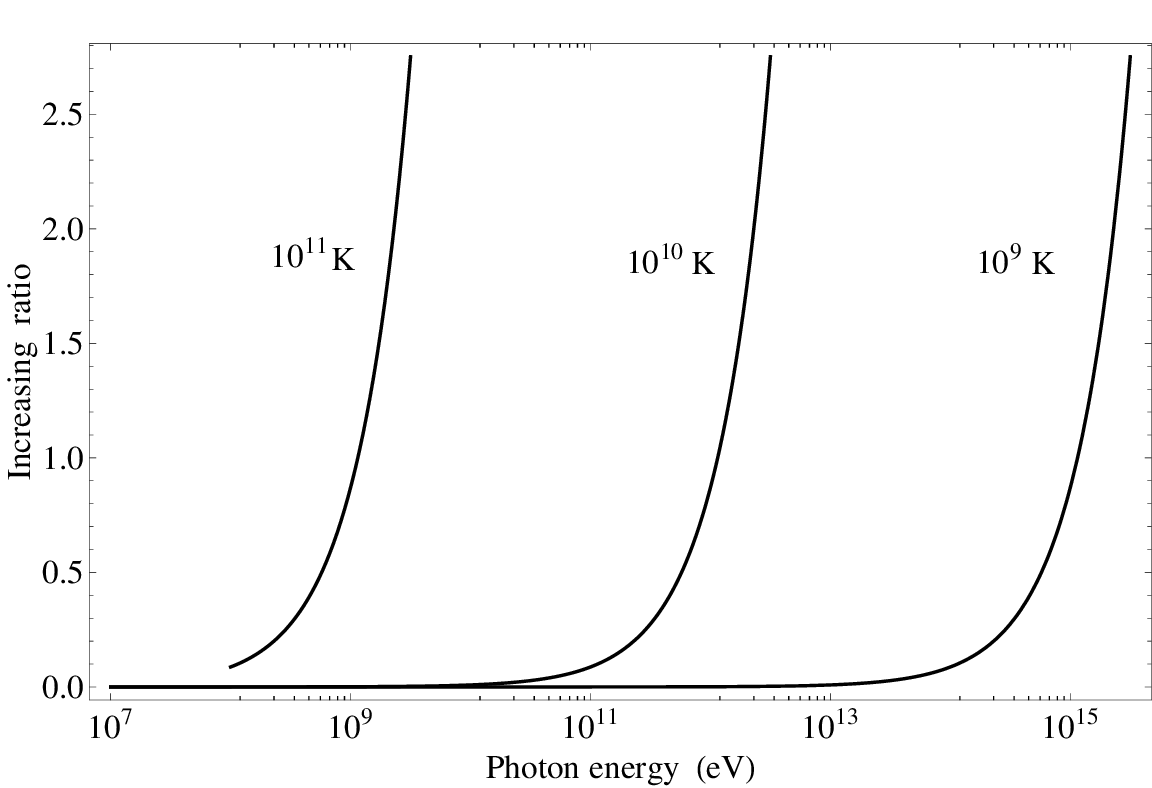}
\caption{The increasing ratio (\ref{23}) of the number of photons versus the photon energy (eV) in the region $\beta \omega \gg 1$ for temperatures $10^{9-11}$~K.}
\end{center}\end{figure}

It would be apposite to check for possible mixing of the effects mentioned here with those coming from including loops in the ordinary electrodynamics. In particular, at energies much less than the electron rest energy, $mc^2$, the dynamics of photons is governed by an effective field theory which is the extension of the Maxwell one by the so-called Euler-Heisenberg Lagrangian \cite{euler,itzy,ravndal},
\begin{equation}
\mathcal{L}_\mathrm{E.H.}=  \frac{8\,\alpha^2}{45\,m^4}\Big(Q^2+\frac{7}{4}P^2\Big),
\end{equation}
in which $Q$ and $P$ are given by (\ref{3}) and (\ref{4}), respectively. This effective theory can be used to find the corrections to the Stefan's law for temperatures $k_\mathrm{B}T\ll mc^2$ \cite{ravndal}
\begin{equation}
\Delta U_\mathrm{E.H.}=\frac{7\cdot 22\;\pi^4\;\alpha^2}{3^5\cdot 5^3}\frac{T^8}{m^4},
\end{equation}
in units with  $c=\hbar=k_\mathrm{B}=1$. So for the ratio of the correction to the total energy one gets
\begin{equation}
\frac{\Delta U_\mathrm{E.H.}}{U_\mathrm{Planck}} \simeq 4\times 10^{-5} \, \left(\frac{T}{m}\right)^4
\end{equation}
which is valid for $T\ll m$, or equivalently $T\ll 6\times 10^9$~K. First, even for $T/m=0.3$ the above ratio is of order $10^{-7}$, and so in the range of validity of the effective theory the corrections are quite small. Second, one mentions that the range of validity of the corrections by the effective theory is different form the range that which the Born-Infeld theory might find for the observational effects. This is simply due to the value of the reference energy much less than that defined by the effective theory, that is the rest energy of the electron \cite{euler,itzy,ravndal}.

In the opposite limit $T\gg m$, one also can find corrections to the black-body radiation of the Maxwell theory. This can be done in the so-called massless fermion limit of the ordinary electrodynamics. The expressions in this limit are already known up to several orders of electron charge, $e$, and are quite different with those coming from the Born-Infeld theory \cite{tggm}.

The other possibility is to look for the implications of the present results on the models for
the extremely hot early universe. As any remarkable change in the energy distribution of the photons in the early universe has direct implications on the evolution of the universe in different eras, it is reasonable to look for the possible consequences of the present results on the early-universe models.

Apart from the problem of infinities which was the original motivation for the suggestion of the Born-Infeld theory, this theory has found a renewal interest since the discovery that
the low-energy effective theories that govern the massless vector modes of open
string theories are of Born-Infeld type. In particular, at tree level, the effective Lagrangian which depends on the Abelian field strength $F_{\mu\nu}$ but not on its derivatives, sum up to the Born-Infeld one \cite{tseytlin}. In this new raise of the theory the parameter $\gamma$ is determined by the corresponding string theory, which is supposedly several orders of magnitude higher than the before-mentioned lower limit. Hence, the implications of this interpretation of the Born-Infeld theory are expected to be considerable only in the first minutes of the early universe, when the temperature has been high enough to be comparable with string theory energy scales. By this way of interpretation, various kinds of indirect tests of string-theory-based models might be provided.

\vspace{0.1cm}
\noindent\textbf{Acknowledgments:}\\
The work by A. H. F. is partially supported by the Research Council of Alzahra University.

\end{document}